\documentclass[11pt,a4paper]{article}    

\usepackage[margin=1in]{geometry} 
\usepackage{amsmath,amssymb,amsthm}
\usepackage{color}
\usepackage{relsize}
\usepackage[perpage,symbol*]{footmisc}
\usepackage{authblk}
\usepackage{enumerate}
\usepackage{cases}
\usepackage{cite}

\newtheorem{theorem}{Theorem}[section]

\newtheorem{lemma}[theorem]{Lemma}

\newtheorem{remark}{Remark}

\newcommand{\be}{\begin{equation}}
\newcommand{\ee}{\end{equation}}
\newcommand{\bea}{\begin{eqnarray}}
\newcommand{\eea}{\end{eqnarray}}
\newcommand{\e}{{\rm e}}

\newtheorem{thm}[theorem]{Theorem}

\numberwithin{equation}{section}

\title{Gap Probability Distribution of Gaussian Unitary Ensembles and Painlev\'{e} V Equation}
\author[1]{Shengjie Zhang}
\author[1,\thanks{Author to whom any correspondence should be addressed. E-mail: lvshulin1989@163.com}]{Shulin Lyu}
\affil[1]{School of Mathematics and Statistics, Qilu University of Technology (Shandong Academy of Sciences), Jinan
250353, China}
\date{\today}

\emergencystretch=\maxdimen
\begin{document}
\maketitle

\begin{abstract}
We consider the Hankel determinant generated by the moments of the even weight function $\e^{-x^2}(A+B\theta(x^2-a^2)), x\in(-\infty,+\infty), a>0, A\ge0, A+B\ge0$. It is intimately related to the gap probability of the Gaussian unitary ensembles on $(-a,a)$ or $(-\infty,-a)\cup(a,+\infty)$. We derive the ladder operators for the monic polynomials orthogonal with respect to this weight function and three supplementary conditions. By using them and differentiating the orthogonality relation, we get difference and Riccati equations for the two auxiliary quantities introduced in the ladder operators.  From these equations, we obtain a second-order difference equation and a second-order second-degree ordinary differential equation satisfied by the coefficient of the three-term recurrence relation for the monic orthogonal polynomials. Moreover, we establish the second-order fourth-degree ordinary differential equation satisfied by the logarithmic derivative of the Hankel determinant. As $n\to\infty$ and $a\to0^{+}$ such that $\tau=2\sqrt{2n}a$ is fixed, this equation is reduced to the $\sigma$-form of a Painlev\'{e} V equation in the variable $\tau$.
\end{abstract}

\noindent

\section{Introduction}

In random matrix theory, the joint probability density function with respect to the eigenvalues of the Gaussian unitary ensemble (GUE for short) is given as follows\cite{22} 
\begin{align*}
	p(x_1, x_2, \cdots, x_n)=\dfrac{1}{C_n}\cdot\dfrac{1}{n!}\prod\limits_{1\le i<j\le n}(x_i-x_j)^2\prod\limits_{l=1}^{n}\e^{-x_l^2}, \quad l=1,2,\cdots,n,
\end{align*}
where $x_l\in(-\infty,+\infty)$ for $l=1,2,\cdots,n$. Here $n!C_n$ is the normalization constant\cite{22}, and 
\begin{align*}
	C_n&=\dfrac{1}{n!}\int_{(-\infty,+\infty)^n}\prod\limits_{1\le i<j\le n}(x_i-x_j)^2\prod\limits_{l=1}^{n}\e^{-x_l^2}\prod\limits_{l=1}^{n}dx_l\\
	&=\mathrm{det}\left(\int_{-\infty}^{+\infty} x^{i+j}\e^{-x^2}dx\right)_{i,j=0}^{n-1}\\
	&=(2\pi)^{n/2}2^{-n^2/2}G(n+1),
\end{align*}
where $G(\cdot)$ is the Barnes-G function defined by $G(n+1)=\Gamma(n)G(n)$ with $G(1)=1$\cite{31}. 
The above second equation can be gotten by using Heine's Identity\cite{28}.
It is easy to show that the probability that the eigenvalues of GUE are within the interval $I\subseteq(-\infty,+\infty)$ is
\begin{align}
	\mathrm{Prob}(x_l\in I, l=1,2,\cdots,n)&=\dfrac{1}{C_n}\cdot\dfrac{1}{n!}\int_{I^n}\prod\limits_{1\le i<j\le n}(x_i-x_j)^2\prod\limits_{l=1}^{n}\e^{-x_l^2}\prod\limits_{l=1}^{n}dx_l\nonumber\\
	&=\dfrac{1}{C_n}\mathrm{det}\left(\int_I x^{i+j}\e^{-x^2}dx\right)_{i,j=0}^{n-1}.\label{001}
\end{align}

Hankel determinants which are used to deal with the probability of the  eigenvalues of Hermitian matrix play an important role in random matrices theory. In this paper, we consider the Hankel determinant generated by the moments of Gaussian weight function with two jump discontinuities at $a$ and $-a$, namely
 \begin{align}
 	D_n(a):&=\mathrm{det}\left(\int_{-\infty}^{+\infty}x^{i+j}w(x;a)dx\right)_{i,j=0}^{n-1},\label{002}
 \end{align}
 where the weight function reads 
 \begin{align*}
 	w(x;a)=w_0(x)w_J(x;a), 
 \end{align*}
 with  
 \begin{align*}
 	w_0(x):=\e^{-x^2},\quad  
 	w_J(x;a):=A+B\theta(x^2-a^2),\quad a>0,\quad x\in(-\infty,+\infty).
 \end{align*}
Here $A$ and $B$ are constants with $A\ge0$ and $A+B\ge0$, and $\theta(\cdot)$ is the Heaviside step function. $\theta(x^2-a^2)$ is 0 for $-a<x<a$ and 1 otherwise. When $A=0$ and $B=1$, combining the equality (\ref{001}) and (\ref{002}), we know that the probability that the interval $(-a,a)$ is free of eigenvalues is 
\begin{align*}
    \mathrm{Prob} \left( x_l\in (-\infty,-a)\cup(a,+\infty), l=1,2,\cdots,n\right) =\dfrac{D_n(a)}{C_n}.
\end{align*}
This case was studied in \cite{05, 20, 22, 29} by the ladder operator approach. Unfortunately, the authors made a mistake during the calculation in \cite{05}. In \cite{20}, the authors have corrected the calculations and obtained the $\sigma$-form of a Painlev\'{e} V equation. When $A=1$ and $B=-1$, $D_n(a)/C_n$ denotes the probability that all eigenvalues lie in the interval $(-a,a)$.

It is easy to show that $D_n(a)$ has the following representation \cite{28}
\begin{align}
	D_n(a)=\prod\limits_{j=0}^{n-1}h_j(a).\label{Dn}
\end{align}
Here we can define the associated monic orthogonal polynomials for the weight function $w(x;a)$ by 
\begin{align}
	\int_{-\infty}^{+\infty}P_j(y;a)P_k(y;a)w(y;a)\,dy=h_j(a)\delta_{jk},\qquad j,k=0,1,2,\cdots,\label{003}
\end{align}
where $P_j(x;a)$ is given by\cite{12}
\begin{align}
	P_j(x;a)=x^j+p(j;a)x^{j-2}+\cdots+P_j(0;a),\qquad n\ge 1.\label{004}
\end{align}
The coefficient $p(n;a)$ of $x^{j-2}$ plays a vital role in subsequent sections. Since the weight function $w(x;a)$ is even, we have  
\begin{align}
	P_n(x;-a)=(-1)^nP_n(x;a).\label{005}
\end{align} 

Substituting (\ref{004}) into (\ref{003}), we have the three-term recurrence relation  for the orthogonal polynomials\cite{28,30}
\begin{align}
	xP_n(x;a)=P_{n+1}(x;a)+\beta_n(a)P_{n-1}(x;a), \label{006}
\end{align}
with the initial conditions 
\begin{align*}
	P_0(x;a)=1,\qquad\qquad   
	\beta_{0}(a)P_{-1}(x;a)=0.
\end{align*}
Setting $j=n-1$, $j=n$ and $j=n+1$ in (\ref{004}), plugging them into (\ref{006}) and comparing the coefficients of $x^{n-1}$ on both sides of the equation, we find 
\begin{align}
	\beta_n(a)&=p(n,a)-p(n+1,a).\label{007}
\end{align}
Multiplying (\ref{006}) by $P_{n-1}(x,a)w(x,a)$, and integrating both sides of the equation with respect to $x$ yields
\begin{align}
	\beta_n(a)=\dfrac{h_n(a)}{h_{n-1}(a)}. \label{008}
\end{align}
Taking a telescopic sum of (\ref{007}), we get 
\begin{align}
	\displaystyle{\sum\limits_{i=0}^{n-1}\beta_j(a)=-p(n;a)}. \label{009}
\end{align}

According to the three-term recurrence relation (\ref{006}), the following Christoffel-Darboux formula is obtained 
\begin{align}
	\sum_{j=0}^{n-1}\frac{P_j(z)P_j(y)}{h_j}=\frac{P_n(z)P_{n-1}(y)-P_{n-1}(z)P_{n}(y)}{h_{n-1}(z-y)}.\label{CDf}
\end{align}
A combination of (\ref{003})-(\ref{006}) and (\ref{008}) gives the following lowering operator for $P_n(z;a)$:
\begin{align*}
	\left(\dfrac{d}{dz}+B_n(z;a) \right)P_n(z;a)&=\beta_n(a)A_n(z;a)P_{n-1}(z;a),
\end{align*}
where $A_n(z;a)$ and $B_n(z;a)$ are closely related to $P_n(z)$. According to the recurrence relation, the Christoffel-Darboux formula and the definitions of $A_n(z;a)$ and $B_n(z;a)$, one obtains the following raising operator
\begin{align*}
	\left(\dfrac{d}{dz}-{\rm v}_0'(z)-B_n(z;a) \right)P_{n-1}(z;a)=-A_{n-1}(z;a)P_n(z;a),
\end{align*}	
where ${\rm v}_0(z):=-\ln w_0(z)$ and three supplementary conditions satisfied by $A_n(z; a)$ and $B_n(z;a)$ which are numbered (\ref{S1}), (\ref{S2}) and (\ref{S2'}). 
For $w(x;a)$, we calculate $A_n(z;a)$ and $B_n(z;a)$, where two auxiliary quantities show up. Substituting the expressions of $A_n(z;a)$ and $B_n(z;a)$ into three supplementary conditions (\ref{S1}), (\ref{S2}) and (\ref{S2'}) gives us a series of difference equations. Differentiating the orthogonality relation (\ref{003}) over $a$, we get differential equations. Finally, by combining the difference equations with the differential equations, the second-order differential equation satisfied by the logarithmic derivative of $D_n(a)$ is deduced in this paper. The above mentioned method is known as the ladder operator approach\cite{09,30}.

The ladder operator approach is an important tool to study unitary ensembles involving one variable, two variables or several variables. For example, in \cite{03}, the authors presented the ladder operator approach adapted to monic orthogonal polynomials with respect to the discontinuous weight function $w(x;t)=w_0(x)(A+B\theta(x-t)), A\ge 0, A+B>0$, where $w_0(x)$ is a generic smooth function and $\theta(\cdot)$ is the Heaviside step function. When $w_0(x)=x^\alpha\e^{-x}, \alpha>0, x\in\left [ 0,+\infty \right )$, the $\sigma$-form of a Painlev\'{e} V equation was derived for the logarithmic derivative of the associated Hankel determinant. In \cite{08}, the authors investigated the Hankel determinant generated by a deformed Laguerre weight function with double-time and derived a second partial differential equation (PDE for short) for the logarithmic derivative of the associated Hankel determinant which was reduced to the $\sigma$-form of a Painlev\'{e} V equation in case that the Hankel determinant has only one variable. The logarithmic derivative of the Hankel determinant of the GUE with $N$ Fisher-Hartwig singularities of root type satisfied the second order PDE in \cite{27}. And for $N=1$, the second order PDE was reduced to the $\sigma$-form of a Painlev\'{e} IV equation. See also \cite{09, 19, 24, 25, 26, 32} for the application of the ladder operator approach to unitary ensembles. 

Using the ladder operator approach, the logarithmic derivative of the Hankel determinant generated by the Laguerre weight function with $m$ jump discontinuities was shown to satisfy a second order PDE\cite{21}, which is reduced to the $\sigma$-form of a Painlev\'{e} V equation when $m=1$. By using the Riemann-Hilbert (RH for short) problem for the associated orthogonal polynomials, the Lax pair has been found. Based on these results, it is shown that the logarithmic derivative of the Hankel determinant satisfies the Hamiltonian of a coupled Painlev\'{e} V system which is a generalization of a Painlev\'{e} V equation. See \cite{10} for the discussion of the Gaussian case. The RH problem for the associated orthogonal polynomials also plays a vital role in random matrices theory. See e.g. \cite{02, 13, 14, 15, 16}.

The gap probability problem for unitary ensembles has attracted a lot of attention. Tracy and Widom investigated the probability that the interval $(-a,a),\;a>0$ is free of eigenvalues in the GUE and the auxiliary quantities associated with the probability satisfied the second order differential equations in \cite{29}. In \cite{04}, via the ladder operator approach, the authors got the probabilities that all the eigenvalues of GUE and the Laguerre unitary ensemble (LUE for short) lie in the interval $(a,b)$. The logarithmic derivatives of the Hankel determinant were shown to satisfy second-order PDEs which can be viewed as two-variable generalization of a Painlev\'{e} IV equation for GUE and a Painlev\'{e} V equation for LUE, respectively. In \cite{01}, through the vertex operator, Adler and van Moerbeke have studied the probabilities of GUE, LUE and Jacobi unitary ensemble (JUE for short) that satisfied nonlinear PDEs which are reduced to Painlev\'{e} IV, V and VI equations, respectively. In \cite{23}, the gap probability that the interval $(-a,a)(0<a<1)$ is free of eigenvalues in JUE was shown to satisfy a second order ordinary differential equation, which was reduced to the $\sigma$-form of a Painlev\'{e} V equation under double scaling limit.

This paper is organized as follows. In Section 2, we derive the ladder operators and three supplementary conditions (\ref{S1}), (\ref{S2}) and (\ref{S2'}) for generic even weight functions $w(x;a)=w_0(x)(A+B\theta(x^2-a^2))$
by using the orthogonality relation. In Section 3, for our weight function $w_0(x)=\e^{-x^{2}}$, by combining  (\ref{S1}) with (\ref{S2'}), we derive the expression of $\beta_{n}(a)$ and a series of difference equations that can be iterated in $n$ for the auxiliary quantities $R_n(a)$ and $r_n(a)$. From this difference equations and the expression of $\beta_{n}(a)$, we find a second order difference equation for $\beta_n(a)$ which can be iterated in $n$. Differentiating the orthogonal relation with respect to $a$, we obtain the differential relations and the Riccati equations satisfied by $R_n(a)$ and $r_n(a)$. According to the derivative of the Riccati equations, we come to the second order differential equations for $\beta_{n}(a)$, $R_n(a)$ and $r_n(a)$. Based on the results in Sections 2 and 3, we get a second order differential equation satisfied by the logarithmic derivative of Hankel determinant. As $n\to\infty$ and $a\to0^{+}$ such that $\tau=2\sqrt{2n}a$ is fixed, this equation is reduced to the $\sigma$-form of a Painlev\'{e} V equation in the variable $\tau$ in Section 4.

\section{Ladder Operators and Supplementary Conditions}

In the following sections, for convenience, we do not display $a$ in $P_n(z;a)$, $w(z;a)$, $h_n(a)$ and $\beta_{n}(a)$ unless necessary. 

In this section, we use the orthogonality relation (\ref{003}) to derive the ladder operators and three supplementary conditions (\ref{S1}), (\ref{S2}) and (\ref{S2'}) for generic even weight functions $w(z)$ defined on $[-\alpha,\alpha],\;\alpha>0$, reading
\begin{align*}
	w(z;a)=w_0(z)w_J(z;a),\;w_J(z;a)=A+B\theta(z^2-a^2),\;A\ge0 ,\; A+B\ge0,
\end{align*} 
where $a<\alpha$, $w_0(z)$ is continuous on $[-\alpha,\alpha]$ and $w_0(-\alpha)=w_0(\alpha)=0$.
For our problem, $w_0(z)=\e^{-z^2},\;\alpha=+\infty$, and for the Jacobi weight which was studied in \cite{23}, $w_0(z)=(1-x^2)^{\beta},\;\beta>0,\;\alpha=1,\;a\in(0,1)$. Note that $\theta(z^2-a^2)=1+\theta(z-a)-\theta(z+a)$, we find 
\begin{align}
	\dfrac{d}{dz}w_J(z;a)&=B(\delta(z-a)-\delta(z+a)),\label{dz}\\
	\dfrac{d}{da}w_J(z;a)&=-B(\delta(z-a)+\delta(z+a)).\label{da}
\end{align} 

\begin{thm} The monic polynomials $\left\lbrace P_{n}(z)\right\rbrace$ orthogonal with respect to $w(z)$ satisfy the lowering operator 
	\begin{align}
		P_n'(z)=-B_n(z)P_n(z)+\beta_nA_n(z)P_{n-1}(z), \label{lo}
	\end{align}
where
\begin{align}
	A_n(z)&:=\dfrac{aR_n(a)}{z^2-a^2}+\dfrac{1}{h_n}\displaystyle{\int_{-\alpha}^{\alpha}\dfrac{{\rm v}_0'(z)-{\rm v}_0'(y)}{z-y}P_n^2(y)w(y)dy},\label{An}\\
	B_n(z)&:=\dfrac{zr_n(a)}{z^2-a^2}+\dfrac{1}{h_{n-1}}\displaystyle{\int_{-\alpha}^{\alpha}\dfrac{{\rm v}_0'(z)-{\rm v}_0'(y)}{z-y}P_n(y)P_{n-1}(y)w(y)dy},\label{Bn}
\end{align}
with ${\rm v}_0(z)=-\ln w_0(z)$ and
\begin{align}
    R_n(a)&:=\dfrac{2BP_n^2(a)w_0(a)}{h_n(a)},\label{Rn}\\
    r_n(a)&:=\dfrac{2BP_n(a)P_{n-1}(a)w_0(a)}{h_{n-1}(a)}.\label{rn}
\end{align}
\end{thm}
\begin{proof}
We start from 
\begin{align}
	P_n'(z)=\displaystyle{\sum\limits_{k=0}^{n-1}C_{n,k}P_k(z)}.\label{010}
\end{align}
According to the 
orthogonality relation (\ref{003}), we know that  
\begin{align}
	C_{n,k}=\dfrac 1 {h_k}\displaystyle{\int_{-\alpha}^{\alpha}P_n'(y)P_k(y)w(y)dy},\quad  k=0,1,\cdots,n-1. \label{011}
\end{align}
Through integration by parts and in view of $w_0(-\alpha)=w_0(\alpha)=0$, we find that
\begin{align}
    C_{n,k}=-\dfrac 1 {h_k}\displaystyle{\int_{-\alpha}^{\alpha} P_n(y)P_k'(y)w(y)dy}-\dfrac 1 {h_k}\displaystyle{\int_{-\alpha}^{\alpha} P_n(y)P_k(y)w'(y)dy}.\label{012}
\end{align}
Since $P_k(y)$ is of degree $k$ with $k\le n-1$, according to the orthogonality relation (\ref{003}), we know that the first integral in the above equation is zero. Hence, noting that $w(z;a)=w_0(z)w_J(z;a)$, (\ref{012}) becomes
\begin{align}
	C_{n,k}=-\dfrac 1 {h_k}\displaystyle{\int_{-\alpha}^{\alpha}P_n(y)P_k(y)w_0'(y)w_J(y)dy}-\dfrac 1 {h_k}\displaystyle{\int_{-\alpha}^{\alpha}P_n(y)P_k(y)w_0(y)w_J'(y)dy}.\label{013}
\end{align}
Since $w_0(y)=\e^{-{\rm v}_0(y)}$, we have $w_0'(y)=-{\rm v}_0'(y)w_0(y)$. Plugging it and (\ref{dz}) into (\ref{013}), we arrive at
\begin{align}
	C_{n,k}=&\dfrac 1 {h_k}\displaystyle{\int_{-\alpha}^{\alpha}P_n(y)P_k(y)w(y){\rm v}_0'(y)dy}-\dfrac 1 {h_k}\displaystyle{\int_{-\alpha}^{\alpha}P_n(y)P_k(y)w_0(y)B(\delta(y-a)-\delta(y+a))dy}\nonumber\\
	=&\dfrac{1}{h_k}\displaystyle{\int_{-\alpha}^{\alpha}P_n(y)P_k(y)w(y)[{\rm v}_0'(y)-{\rm v}_0'(z)]dy}-\dfrac{BP_n(a)P_k(a)w_0(a)}{h_k}\nonumber\\
	&+\dfrac{BP_n(-a)P_k(-a)w_0(-a)}{h_k},\label{014}
\end{align}
where to obtain the second identity, we make use of the orthogonality relation. 

Inserting (\ref{014}) into (\ref{010}) gives 
\begin{align}
P_n'(z)=&\displaystyle{\int_{-\alpha}^{\alpha}\sum\limits_{k=0}^{n-1}\left(\dfrac{P_k(z)P_k(y)}{h_k} \right) P_n(y)w(y)[{\rm v}_0'(y)-{\rm v}_0'(z)]\,dy}-\sum\limits_{k=0}^{n-1}\left(\dfrac{P_k(z)P_k(a)}{h_k} \right) BP_n(a)w_0(a)\nonumber\\
&+\sum\limits_{k=0}^{n-1}\left(\dfrac{P_k(z)P_k(-a)}{h_k} \right)BP_n(-a)w_0(-a). \label{015}
\end{align}
According to the Christoffel-Darboux formula (\ref{CDf}) and the relation (\ref{005}) i.e. $P_n(x;-a)=(-1)^nP_n(x;a)$, we obtain from (\ref{015})
\begin{align*}	
    P_n'(z)=&\displaystyle{\int_{-\alpha}^{\alpha}\dfrac{P_n(z)P_{n-1}(y)-P_{n-1}(z)P_{n}(y)}{h_{n-1}(z-y)}P_n(y)w(y)[{\rm v}_0'(y)-{\rm v}_0'(z)]\,dy}\\
    &+\dfrac{BP_{n}^2(a)w_0(a)P_{n-1}(z)}{h_{n-1}(z-a)}-\dfrac{BP_{n}(a)P_{n-1}(a)w_0(a)P_{n}(z)}{h_{n-1}(z-a)}\\
    &-\dfrac{BP_{n}^2(a)w_0(a)P_{n-1}(z)}{h_{n-1}(z+a)}-\dfrac{BP_{n}(a)P_{n-1}(a)w_0(a)P_{n}(z)}{h_{n-1}(z+a)}
\end{align*}
\begin{align*}	
	=&-\left[\displaystyle{\dfrac{1}{h_{n-1}}\int_{-\alpha}^{\alpha}\dfrac{{\rm v}_0'(z)-{\rm v}_0'(y)}{z-y}P_n(y)P_{n-1}(y)w(y)\,dy}+\dfrac{2zBP_n(a)P_{n-1}(a)w_0(a)}{h_{n-1}(z^2-a^2)}\right] P_n(z)\\
	&+\beta_n\left[ \displaystyle{\dfrac{1}{h_n}\int_{-\alpha}^{\alpha}\dfrac{{\rm v}_0'(z)-{\rm v}_0'(y)}{z-y}P_n^2(y)w(y)\,dy}+\dfrac{2aBP_n(a)P_{n-1}(a)w_0(a)}{h_{n}(z^2-a^2)}\right] P_{n-1}(z).
\end{align*}
Finally, we get the lowering operator (\ref{lo}).
\end{proof}
Basing on the definition of $A_n(z)$ and $B_n(z)$, the recurrence relation (\ref{006}) and the orthogonality relation (\ref{003}), ones will obtain (\ref{S1}) in the following Lemma.
\begin{lemma} The quantities $A_n(z)$ and $B_n(z)$ satisfy the following identity:
	\begin{align}
		B_n(z)+B_{n+1}(z)=zA_n(z)-{\rm v}_0'(z),\quad n\ge0. \tag{$S_1$}\label{S1}
	\end{align}
	Here ${\rm v}_0(z)=-\ln w_0(z)$.
\end{lemma}

\begin{proof}
	According to the definitions of $B_n(z)$ and $r_n(a)$ given by \eqref{Bn} and \eqref{rn}, we find 
\begin{align*}
	B_n(z)+B_{n+1}(z)=&\dfrac{2zBP_n(a)w_0(a)\left[ \beta_nP_{n-1}(a)+P_{n+1}(a)\right] }{h_{n}(z^2-a^2)}\nonumber\\
	&+\dfrac{1}{h_{n}}\displaystyle{\int_{-\alpha}^{\alpha} \dfrac{{\rm v}_0'(z)-{\rm v}_0'(y)}{z-y}P_n(y)[\beta_nP_{n-1}(y)+P_{n+1}(y)]w(y)dy},
\end{align*} 
where we make use of the fact that $\beta_n=h_{n}/h_{n-1}$.
Noting that the quantity $\beta_nP_{n-1}+P_{n+1}$ appears in both terms on the right hand side of the above equation, we make use of the recurrence relation (\ref{006}) to get rid of it and get the following equation 
\begin{align}
	&B_n(z)+B_{n+1}(z)=\dfrac{2aBP^2_n(a)w_0(a)\cdot z}{h_{n}(z^2-a^2)}+\dfrac{1}{h_{n}}\displaystyle{\int_{-\alpha}^{\alpha} \dfrac{{\rm v}_0'(z)-{\rm v}_0'(y)}{z-y} yP_n^2(y)w(y)dy}\nonumber\\
	&=\dfrac{aR_n\cdot z}{z^2-a^2}+\dfrac{1}{h_{n}}\displaystyle{\int_{-\alpha}^{\alpha}  \dfrac{{\rm v}_0'(z)-{\rm v}_0'(y)}{z-y}zP_n^2(y)w(y)dy}-{\rm v}_0^{'}(z)+\dfrac{1}{h_{n}}\displaystyle{\int_{-\alpha}^{\alpha} {\rm v}_0'(y)P_n^2(y)w(y)dy},\label{016}
\end{align}
where by using the definition of (\ref{Rn}) and the identity $y/(z-y)=z/(z-y)-1$, we get (\ref{016}).

Now we look at the last integral in (\ref{016}). According to ${\rm v}_0(y)=-\ln w_0(y)$, we have ${\rm v}_0'(y)=-w_0'(y)/w_0(y)$. Hence, this integral becomes
\begin{align}
\dfrac{1}{h_{n}}\displaystyle{\int_{-\alpha}^{\alpha} {\rm v}_0'(y)P_n^2(y)w(y)dy}=-\dfrac{1}{h_{n}}\displaystyle{\int_{-\alpha}^{\alpha}w_0'(y)P_n^2(y)w_J(y)dy}.\label{017}
\end{align}
Through integration by parts, in view of $w_0(-\alpha)=w_0(\alpha)=0$ and the orthogonality relation (\ref{003}), the integral in the right hand side of (\ref{017}) is shown to be $0$. Hence, the last integral in (\ref{016}) is zero and consequently (\ref{016}) leads us to (\ref{S1}).
\end{proof}

Using the recurrence relation (\ref{006}) to get rid of $\beta_{n}P_{n-1}(z)$ in (\ref{lo}), we get
\begin{align*}
	P_n'(z)=-B_n(z)P_n(z)+zA_n(z)P_n(z)-A_n(z)P_{n+1}(z).
\end{align*}
Eliminating $B_n(z)$ in the above equation by using \eqref{S1}, we find 
$$P_{n}'(z)=\left( {\rm v}_0'(z)+B_{n+1}(z)\right) P_{n}(z)-A_{n}(z)P_{n+1}(z).$$
Replacing $n$ by $n-1$ in the above equation, we get the following raising operator.
\begin{thm} The monic orthogonal polynomials $\left\lbrace P_{n}(z)\right\rbrace $ satisfy the raising operator 
	\begin{align}
		\left(\dfrac{d}{dz}-{\rm v}_0'(z)-B_n(z) \right)P_{n-1}(z)=-A_{n-1}(z)P_n(z).\label{ro}
	\end{align}	
\end{thm}
\begin{remark}
	Combining \eqref{lo} with \eqref{ro} to eliminate $P_{n-1}(z)$ in \eqref{ro}, we get the following second order ordinary differential equation for $P_{n}(z)$:
	\begin{align}
		P''_n(z)+Q_n(z)P'_n(z)+S_n(z)P_n(z)=0,\label{heun}
	\end{align}
	where
	\begin{align*}
		Q_n(z)&=-\dfrac{A'_n(z)}{A_n(z)}-{\rm v}_0'(z),\\
	    S_n(z)=B'_n(z)-\dfrac{A'_n(z)}{A_n(z)}B_n(z)-&{\rm v}_0'(z)B_n(z)-B^2_n(z)+\beta_nA_n(z)A_{n-1}(z).
	\end{align*}
	
	Here $\left\lbrace P_{n}(z)\right\rbrace $ are orthogonal with respect to $w(z;a)=w_0(z)\left(A+B\theta(z^2-a^2) \right) , z\in[-\alpha,\alpha]$, where $A\ge0 ,\; A+B\ge0$, $a<\alpha$, $\alpha>0$,  $w_0(z)$ is continuous and $w_0(-\alpha)=w_0(\alpha)=0$. When $A=0$, $B=1$, $w_0(z)=\e^{-z^2}$ or $w_0(z)=(1-z^2)^{\beta}$, \eqref{heun} was shown to be connected with Heun equations\cite{07}. Below are the results: 
	
	For the Gaussian case, i.e. $w(z;a)=\e^{-z^2}\theta(z^2-a^2), z\in(-\infty,+\infty), a>0$, by introducing a new variable $u=z^2/a^2$ and setting $t=a^2$, as $n\to\infty$, \eqref{heun} was shown to satisfy the confluent Heun equation in the variable $u$ (see \cite[(4.4)]{07}).
	For the Jacobi case, i.e. $w(z;a)=(1-z^2)^{\beta}\theta(z^2-a^2),\;\beta>0,\; z\in[-1,1],\; 0<a<1$,  by introducing a new variable $u=z^2$ and setting $t=a^2$,  in the variable $u$,  \eqref{heun} has been reduced to the general Heun equation (see \cite[(4.14)]{07}).
	
	Note that \eqref{heun} is determined by $Q_n(z)$ and $S_n(z)$ which depend on $A_n(z)$, $B_n(z)$, $\beta_n$ and ${\rm v}_0'(z)$. For our problem, we observe that $A$ and $B$ do not appear in $A_n(z)$ and $B_n(z)$. Actually only $B$ is included in the definitions of the auxiliary quantities. Moreover, we find that the expressions of $A_n(z)$ and $B_n(z)$ given by \eqref{301} and \eqref{302} have the same forms with the ones in \cite{07} where $A=0$ and $B=1$. Hence, we conclude that, as $n\to\infty$, \eqref{heun} for our problem can also be transformed into the confluent Heun equation given in \cite{07} (see (4.4) therein) .
\end{remark}

Next, we use the ladder operators and (\ref{S1}) to deduce another two identities for $A_n(z)$ and $B_n(z)$ in the following lemma.
\begin{lemma} We have
	\begin{align}
		1+z\left(B_{n+1}(z)-B_n(z) \right)=\beta_{n+1}A_{n+1}(z)-\beta_nA_{n-1}(z),\tag{$S_2$}\label{S2}\\
		\sum\limits_{j=0}^{n-1} A_j(z)+B_n^2(z)+B_n(z){\rm v}_0'(z)=\beta_nA_n(z)A_{n-1}(z). \tag{$S_2'$}\label{S2'}
	\end{align}
\end{lemma}
\begin{proof}
	We first prove \eqref{S2}. Taking derivative on both sides of the recurrence relation \eqref{006} with respect to $z$ gives us
	\begin{align*}
	P_n(z)+zP_n'(z)=P_{n+1}'(z)+\beta_nP_{n-1}'(z).
    \end{align*}
    Using the lowering operator \eqref{lo} to get rid of $P_{n+1}'(z)$ and $P_n'(z)$ in the above equation, and removing $P_{n-1}'(z)$ by making use of the raising operator \eqref{ro}, we come to  
    \begin{align*}
    	-B_{n+1}(z)P_{n+1}(z)+&\left[B_{n}(z)+{\rm v}_0'(z)-zA_{n}(z) \right]\beta_nP_{n-1}(z)\nonumber\\
    	&=\left[1-zB_{n}(z)-\beta_{n+1}A_{n+1}(z)+\beta_{n}A_{n-1}(z)\right]P_n(z).
    \end{align*}
	Replacing $P_{n+1}(z)$ in the above identity by using the recurrence relation (\ref{006}), we are led to  
    \begin{align}
    	\left[1+z(B_{n+1}(z)-B_{n}(z))-\beta_{n+1}A_{n+1}(z)+\beta_{n}A_{n-1}(z)\right]P_n(z)\nonumber\\
    	=\left[B_{n}(z)+B_{n+1}(z)+{\rm v}_0'(z)-zA_{n}(z) \right]\beta_nP_{n-1}(z).\label{018}
    \end{align}
    According to (\ref{S1}), we find that the right hand side of (\ref{018}) is zero. Hence we arrive at (\ref{S2}).
	
	Multiplying both sides of (\ref{S2}) by $A_{n}(z)$ leads us to 
	\begin{align}
		A_{n}(z)+(zA_{n}(z))\cdot(B_{n+1}(z)-B_n(z))=\beta_{n+1}A_{n+1}(z)A_{n}(z)-\beta_nA_{n}(z)A_{n-1}(z).\label{023}
	\end{align}
	Removing $zA_{n}(z)$ by using (\ref{S1}), we get 
	\begin{align*}
		A_{n}(z)+{\rm v}_0'(z)(B_{n+1}(z)-B_n(z))+(B_{n+1}^2(z)-B_n^2(z))=\beta_{n+1}A_{n+1}(z)A_{n}(z)-\beta_nA_{n}(z)A_{n-1}(z).
	\end{align*}
	Replacing $n$ by $j$ in this equation and summing both sides of it from $j=0$ to $j=n-1$, in view of $A_{-1}(z)=B_0(z)=0$, we obtain (\ref{S2'}). 
\end{proof}
\begin{remark}
	The derivation of the three supplementary conditions \eqref{S1}, \eqref{S2} and \eqref{S2'} is also presented in \cite{09,10,27,30}. However, the weight function in our problem is even whose associated orthogonal polynomials have obviously different properties compared with those in \cite{09,10,27}.
	
	In \cite{09}, the ladder operators and the three compatibility conditions were derived for orthogonal polynomials with respect to the weight function whose potential function is twice continuously differentiable and convex. The Gaussian weight function with Fisher-Hartwig singularities of jump type \cite{10} and root type \cite{27} was discussed.
	
	In \cite{30}, the ladder operators were derived by using the Riemann-Hilbert problem for orthogonal polynomials, from which \eqref{S1} was deduced.
\end{remark}
\section{Difference Equations, Riccati Equations and Second Order Differential Equations}
This section is divided into two parts. In the first part, we get the expressions for $A_n(z)$ and $B_n(z)$ in terms of two auxiliary quantities $R_n(a)$ and $r_n(a)$. Substituting them into \eqref{S1} and \eqref{S2'}, we obtain the expression of $\beta_{n}$ and $p(n,a)$, the coefficient of $z^{n-2}$ in the orthogonal polynomial of degree $n$, in terms of $R_n$ and $r_n$ which satisfy a system of difference equations. With these results, we establish a second order difference equation for $\beta_{n}$ which can be iterated in $n$. In the second part, by differentiating the orthogonality relation with respect to $a$, we get the differential relations between the derivatives of $\left\lbrace \ln h_n,\,p(n,a)\right\rbrace $ and $\left\lbrace R_n,\, r_n\right\rbrace $. Combining them with the expressions of $\beta_{n}$ and $p(n,a)$, we come to two Riccati equations for $R_n$ and $r_n$, from which we get the second order differential equations satisfied by $\beta_{n}$, $R_n$ and $r_n$.

For convenience, unless necessary we will not display $a$ in $\beta_{n}(a)$,  $r_n(a)$ and $R_n(a)$ in the following sections. 
\subsection{Difference Equations}
For our problem, $w_0(z)=\e^{-z^{2}},\;z\in(-\infty,+\infty)$, we have ${\rm v}_0(z)=-\ln w_0(z)=z^2$, so that ${\rm v}'_0(z)=2z$. Plugging it into (\ref{An}) and (\ref{Bn}), we obtain the following expressions of $A_n(z)$ and $B_n(z)$ in terms of $R_n(a)$ and $r_n(a)$.
\begin{lemma}For our problem, we have 
    \begin{align}
		A_n(z)&=\dfrac{aR_n(a)}{z^2-a^2}+2,\label{301}\\
		B_n(z)&=\dfrac{zr_n(a)}{z^2-a^2},\label{302}
	\end{align}
where the auxiliary quantities $R_n$ and $r_n$ are given by (\ref{Rn}) and (\ref{rn}), reading
	\begin{align}
		R_n(a)&=\dfrac{2BP_n^2(a)\e^{-a^{2}}}{h_n(a)},\label{303}\\
		r_n(a)&=\dfrac{2BP_n(a)P_{n-1}(a)\e^{-a^{2}}}{h_{n-1}(a)}.\label{304}
	\end{align}
\end{lemma}

First, plugging (\ref{301}) and (\ref{302}) into (\ref{S1}), we get 
\begin{align}
	\dfrac{z(r_n+r_{n+1})}{z^2-a^2}=\dfrac{z\cdot aR_n}{z^2-a^2}.\label{305}
\end{align}
Multiplying both sides of \eqref{305} by $z^2-a^2$ and letting $z\to a$, we obtain the difference equation
\begin{align}
	r_n+r_{n+1}&=aR_n.\label{d}
\end{align}
Substituting (\ref{301}) and (\ref{302}) into (\ref{S2'}), we arrive at
\begin{align}
	\sum\limits_{j=0}^{n-1}\left(\dfrac{aR_j}{z^2-a^2}+2\right)+\dfrac{z^{2}r^2_n}{(z^2-a^2)^2}&+\dfrac{2z^{2}r_n}{z^2-a^2}=4\beta_{n}+\dfrac{2a\beta_{n}(R_n+R_{n-1})}{z^2-a^2}+\dfrac{a^{2}\beta_{n}R_nR_{n-1}}{(z^2-a^2)^2}.\label{306}
\end{align}
Since 
\begin{align*}
	\dfrac{z^2}{z^2-a^2}&=1+\dfrac{a^2}{z^2-a^2},\\
	\dfrac{z^2}{(z^2-a^2)^2}&=\dfrac{1}{z^2-a^2}+\dfrac{a^2}{(z^2-a^2)^2},
\end{align*}
equality (\ref{306}) is reduced to 
\begin{align}
	2n+2r_n+\dfrac{a\sum\limits_{j=0}^{n-1}R_j+r_n^2+2a^{2}r_n}{z^2-a^2}+\dfrac{a^{2}r^2_n}{(z^2-a^2)^2}=&4\beta_{n}+\dfrac{2a\beta_{n}(R_n+R_{n-1})}{z^2-a^2}\nonumber\\
	&+\dfrac{a^{2}\beta_{n}R_nR_{n-1}}{(z^2-a^2)^2}.\label{307}
\end{align}
Multiplying both sides of \eqref{307} by $(z^2-a^2)^2$ and letting $z\to a$, we get
\begin{align}
	r^2_n=\beta_{n}R_nR_{n-1},\label{c}
\end{align}
and consequently, \eqref{307} becomes 
\begin{align}
	2n+2r_n+\dfrac{a\sum\limits_{j=0}^{n-1}R_j+r_n^2+2a^{2}r_n}{z^2-a^2}=4\beta_{n}+\dfrac{2a\beta_{n}(R_n+R_{n-1})}{z^2-a^2}.\label{313}
\end{align}
Multiplying both sides of \eqref{313} by $z^2-a^2$ and letting $z\to a$, we find
\begin{align}
	a\sum\limits_{j=0}^{n-1}R_j+r_n^2+2a^{2}r_n=2a\beta_{n}(R_n+R_{n-1}).\label{b}
\end{align}
Hence, \eqref{313} now reads 
\begin{align}
	2\beta_{n}=n+r_n.\label{a}
\end{align}

Combining this equation with \eqref{d} and \eqref{c}, we come to the second order difference equation for $\beta_n$.
\begin{thm}The recurrence coefficient $\beta_n$ is shown to satisfy the following second order difference equation
	\begin{align*}
		a^2\left(2\beta_{n}-n \right)^2=\beta_{n}\left(2\beta_{n}+2\beta_{n+1}-2n-1 \right)\left(2\beta_{n}+2\beta_{n-1}-2n+1 \right),
	\end{align*}
	which can be iterated in $n$ with the initial conditions
	\begin{align*}
		\beta_{0}=0, \quad  \beta_{1}=\dfrac{1}{2}+\dfrac{B\cdot a\e^{-a^2}}{\int_{-\infty}^{+\infty}\e^{-x^2}(A+B\theta(x^2-a^2))dx}.
	\end{align*}
\end{thm}
\begin{proof}
	Inserting \eqref{a} into \eqref{c}, we have 
	\begin{align*}
		2r_n^2=\left(r_n+n \right)R_nR_{n-1}.
	\end{align*}
	Using \eqref{d} to remove $R_n$ and $R_{n-1}$ in the above equation, we get the following difference equation for $r_n$
	\begin{align*}
		2a^2r_n^2=\left(r_n+n \right) \left(r_{n}+r_{n+1} \right)\left(r_{n}+r_{n-1} \right), 
	\end{align*}
	which, according to \eqref{a}, gives us the desired difference equation for $\beta_n$.
	
	Note that we have the following initial conditions
	\begin{align*}
		r_0=0,\qquad
		R_0=\dfrac{2B\e^{-a^2}}{\displaystyle{\int_{-\infty}^{+\infty}\e^{-x^2}(A+B\theta(x^2-a^2))dx}}.
	\end{align*}
	Hence, from \eqref{a} with $n=0$, it follows that $\beta_{0}=0$. In addition, 
	when $n=0$, \eqref{d} is reduced to $aR_0=r_{1}$. Plugging it into \eqref{a}, we come to the value of $\beta_{1}$.
\end{proof}
Now, we focus on $p(n,a)$ which is the coefficient of $z^{n-2}$ in the monic orthogonal polynomial of degree $n$. It plays a crucial role for the derivation of the PDE satisfied by the logarithmic derivative of $D_n(a)$ in the following section. By using \eqref{d} and (\ref{c})-(\ref{a}), the expression of $p(n,a)$ in terms of the auxiliary quantities $r_n$ and $R_n$ is deduced as follows.
\begin{lemma}
	The quantity $p(n,a)$ is expressed in terms of the auxiliary quantities $r_n$ and $R_n$ by 
	\begin{align}
		p(n,a)=-\dfrac{a}{4}(n+r_n)R_n-\dfrac{ar^2_n}{2R_n}+\dfrac{r^2_n}{4}+\dfrac{a^2r_n}{2}+\dfrac{r_n}{4}-\dfrac{n^2-n}{4}. \label{308}
	\end{align}
\end{lemma}
\begin{proof}
		Eliminating $R_{n-1}$ and $\beta_{n}$ in \eqref{b} by using \eqref{c} and \eqref{a} leads us to 
	\begin{align}
		a\sum\limits_{j=0}^{n-1}R_j=a(n+r_n)R_n+\dfrac{2ar^2_n}{R_n}-r^2_n-2a^2r_n. \label{309}
	\end{align}
	A telescopic sum of (\ref{d}) gives us
	\begin{align}
		a\sum\limits_{j=0}^{n-1}R_{j}=2\sum\limits_{j=0}^{n-1}r_j+r_n. \label{310}
	\end{align}
	Combining \eqref{309} and \eqref{310} to remove $a\sum\limits_{j=0}^{n-1}R_{j}$, we get 
	\begin{align}
		\sum\limits_{j=0}^{n-1}r_{j}=\dfrac{a}{2}(n+r_n)R_n+\dfrac{ar^2_n}{R_n}-\dfrac{r^2_n}{2}-a^2r_n-\dfrac{r_n}{2}. \label{311}
	\end{align}
	Replacing $n$ by $j$ in (\ref{a}), and taking a summation of it from $j=0$ to $j=n-1$, we have  
	\begin{align*}
	    \sum\limits_{j=0}^{n-1}\beta_j&=\dfrac{1}{2}\sum\limits_{j=0}^{n-1}r_{j}+\dfrac{n^2-n}{4}.
	\end{align*}
	Combining it with (\ref{009}) yields  
	\begin{align}
		p(n,a)=-\dfrac{1}{2}\sum\limits_{j=0}^{n-1}r_{j}-\dfrac{n^2-n}{4}.\label{314}
	\end{align}
	Plugging \eqref{311} into the above equation, we get the desired result.
\end{proof}

\subsection{Riccati Equations and Second Order Differential Equations}
In order to derive the Riccati equations for  $R_n$ and $r_n$, we need the following differential relations obtained by differentiating the orthogonality relation (\ref{003}) with respect to $a$. 
\begin{lemma} The quantities $\ln h_n(a)$ and $p(n,a)$ satisfy the following differential relations
	\begin{align}
		\dfrac{d}{da}\ln h_n&=-R_n,\label{312}\\
		\dfrac{d}{da}p(n,a)&=ar_n-\beta_{n}R_n.\label{p'}
	\end{align}
\end{lemma}
\begin{proof}
	Differentiating both sides of
	\begin{align*}
		h_n=\displaystyle{\int_{-\infty}^{\infty}P^2_n(y;a)w(y;a)dy},
	\end{align*} with respect to $a$ gives us
	\begin{align}
		\dfrac{d}{da}h_n=\displaystyle{\int_{-\infty}^{\infty}2P_{n}(y;a)\left( \dfrac{d}{da}P_n(y;a)\right) w(y;a)\,dy}+\displaystyle{\int_{-\infty}^{\infty}P^2_n(y;a)w_0(y)\left( \dfrac{d}{da}w_J(y;a)\right)dy}.\label{318}
	\end{align}
	The first integral in \eqref{318} is zero due to the following identity
	\begin{align}
		\dfrac{d}{da}P_n(y;a)=\left( \dfrac{d}{da}p(n,a)\right)y^{n-2}+\cdots=\left( \dfrac{d}{da}p(n,a)\right)P_{n-2}(y;a)+\cdots, \label{315}
	\end{align}
	and the orthogonality relation \eqref{003}.
	Noting that 
	\begin{align*}
		\dfrac{d}{dt}\theta(x-t)=-\delta(x-t),\qquad  \displaystyle{\int_{-\infty}^{\infty}f(x)\delta(x-t)dx}=f(t).
	\end{align*}
	Using \eqref{da} and \eqref{005}, we find that the second integral in \eqref{318} becomes
	\begin{align*}
		\displaystyle{\int_{-\infty}^{\infty}P^2_n(y;a)w_0(y)\left( \dfrac{d}{da}w_J(y;a)\right)dy}=-2BP^2_n(a)w_0(a).
	\end{align*}
	Finally, we get the following identity from \eqref{318} by making use of the definition of $R_n(a)$ given by \eqref{303}
	\begin{align*}
		h'_n=-R_nh_n.
	\end{align*}
	According to the resulting equation, we get \eqref{312}. 
	
	Similarly, differentiating both sides of 
	\begin{align*}
		\displaystyle{\int_{-\infty}^{\infty}P_n(y;a)P_{n-2}(y;a)w(y;a)\,dy}=0,
	\end{align*} with respect to $a$ leads us to
	\begin{align}
		0=&\displaystyle{\int_{-\infty}^{\infty}\left( \dfrac{d}{da}P_n(y;a)\right) P_{n-2}(y;a)w(y;a)dy}+\displaystyle{\int_{-\infty}^{\infty}P_n(y;a)\left(\dfrac{d}{da} P_{n-2}(y;a)\right)w(y;a)dy}\nonumber\\
		&+\displaystyle{\int_{-\infty}^{\infty}P_n(y;a)P_{n-2}(y;a)w_0(y)\left( \dfrac{d}{da}w_J(y;a)\right) dy}. \label{316}
	\end{align}
	Since $\dfrac{d}{da} P_{n-2}(y;a)$ is of degree $n-4$, the second integral in the right hand side of \eqref{316} is zero by using the orthogonality relation \eqref{003}. 
	Plugging \eqref{315} and \eqref{da} into \eqref{316} to get rid of $\dfrac{d}{da}P_n(y;a)$ and $\dfrac{d}{da}w_J(y;a)$, with the aid of \eqref{005}, we get  
	\begin{align}
		\dfrac{d}{da}p(n,a)=\dfrac{2BP_n(a)P_{n-2}(a)w_0(a)}{h_{n-2}}. \label{317}
	\end{align}
	Replacing $n$, $x$ by $n-1$ and $a$ respectively in (\ref{006}), we have
	\begin{align*}
		P_{n-2}(a)=\dfrac{1}{\beta_{n-1}}\left(aP_{n-1}(a)-P_{n}(a) \right)=\dfrac{h_{n-2}}{h_{n-1}}\cdot\left(aP_{n-1}(a)-P_{n}(a) \right).
	\end{align*}
	Substituting it into (\ref{317}) to remove $P_{n-2}(a)$,  we find 
	\begin{align*}
		\dfrac{d}{da}p(n,a)&=a\cdot\dfrac{2BP_n(a)P_{n-1}(a)w_0(a)}{h_{n-1}}-\dfrac{h_n}{h_{n-1}}\cdot\dfrac{2BP^2_n(a)w_0(a)}{h_{n}}\\
		&=ar_n-\beta_{n}R_n,
	\end{align*}
	where to deduce the second equality we make use of \eqref{008} and \eqref{303}-\eqref{304}.
	Finally, we come to the desired identity.
\end{proof}

Next, we combine the differential relations \eqref{312}-\eqref{p'} with the expressions of $\beta_n(a)$ and $p(n,a)$ which are given by \eqref{008} and \eqref{308}, and the difference equations obtained in the previous subsection to derive the following Riccati equations for $R_n(a)$ and $r_n(a)$.
\begin{lemma} The auxiliary quantities $R_n$ and $r_n$ satisfy the Riccati equations
	\begin{align}
		\dfrac{d}{da}r_n&=\dfrac{2r^2_n}{R_n}-(n+r_n)R_n,\label{eqrn}\\
		\dfrac{d}{da}R_n&=R^2_n+4r_n-2aR_n-\dfrac{2r_nR_n}{a}.\label{eqRn}
	\end{align}
\end{lemma}
\begin{proof}
	Since $\beta_{n}=h_n/h_{n-1}$, we find from \eqref{312} that  
	\begin{align*} 
		\dfrac{d}{da}\beta_n=\beta_n(R_{n-1}-R_{n}).
	\end{align*}
	Substituting (\ref{c}) into the above equation to eliminate $R_{n-1}$, we obtain 
	\begin{align}
		\dfrac{d}{da}\beta_n=\dfrac{r^2_n}{R_n}-\beta_nR_{n}. \label{319}
	\end{align}
	Taking the derivative of (\ref{a}), we have
	\begin{align*} 
		\dfrac{d}{da}\beta_n=\dfrac{1}{2}\dfrac{d}{da}r_n.
	\end{align*} Combining it with (\ref{319}), we come to the Riccati equation for $r_n(a)$,  i.e.\eqref{eqrn}.
	
	Differentiating both sides of (\ref{308}) with respect to $a$ gives us
	\begin{align*}
		\dfrac{d}{da}p(n,a)=&\left( -\dfrac{a}{4}(n+r_n)+\dfrac{ar^2_n}{2R^2_n}\right) \dfrac{d}{da}R_n+\left( \dfrac{a^2+r_n}{2}-\dfrac{aR_n}{4}-\dfrac{ar_n}{R_n}+\dfrac{1}{4}\right) \dfrac{d}{da}r_n\nonumber\\
		&-\dfrac{R_n}{4}(n+r_n)-\dfrac{r^2_n}{2R_n}+ar_n.
	\end{align*}
	Substituting (\ref{p'}) and (\ref{eqrn}) into the above equation to remove $\dfrac{d}{da}p(n,a)$ and $\dfrac{d}{da}r_n$, we get the Riccati equation for $R_n(a)$ (\ref{eqRn}).
\end{proof}
Solving $r_n$$\left( resp.R_n\right)$ from \eqref{eqRn}$\left( resp.\eqref{eqrn}\right)$ and substituting it into \eqref{eqrn}$\left( resp.\eqref{eqRn}\right)$, we obtained the second order ODE satisfied by $R_n$$\left( resp.r_n\right)$. By substituting \eqref{a} into the second order ODE satisfied by
 $r_n$, we have the second order ODE for $\beta_{n}$.
\begin{thm} The auxiliary quantities $\left\lbrace R_n, r_n\right\rbrace $ and the recurrence coefficient $\beta_{n}$ satisfy the following second order differential equations
	\begin{align}
		R''_n=\dfrac{a-R_n}{(2a-R_n)R_n}(R'_n)^2+\dfrac{R_n}{a(2a-R_n)}R'_n&+\dfrac{1}{a}(2a-R_n)(a^2-2n-1-aR_n)R_n,\label{eq048}\\
		4(a^2+r_n)^2\left( (r'_n)^2+8r^{2}_n(n+r_n)\right)&-a^2(r''_n+8nr_n+12r^2_n)^2=0, \label{eq053}\\
		4\left( 2\beta_{n}+a^2-n\right)^2\left((\beta'_{n})^2+4\beta_{n}(2\beta_{n}-n)^2 \right)&-a^2\left(\beta''_{n}+2(2\beta_{n}-n)(6\beta_{n}-n) \right)^2=0.\label{102}
	\end{align}
\end{thm}
\begin{proof}
	Solving $r_n$ from \eqref{eqRn}, we get  
	\begin{align}
		r_n=\dfrac{aR'_n}{2(2a-R_n)}+\dfrac{aR_n}{2}. \label{eq049}
	\end{align}
	Substituting it into \eqref{eqrn}, we are led to \eqref{eq048}.
	
	Multiplying both side of (\ref{eqrn}) by $R_n$, we have
	\begin{align*}
		(n+r_n)R^2_n+r'_nR_n-2r^{2}_n=0.
	\end{align*}
	Solving $R_n$ from this equation, we find 
	\begin{align}
		R_n=\dfrac{-r'_n \pm \sqrt{(r'_n)^2+8r^{2}_n(n+r_n)}}{2(n+r_n)}.\label{320}
	\end{align}
	Substituting the above solution into (\ref{eqRn}), by clearing the square root, we come to \eqref{eq053}.
	By plugging \eqref{a} into \eqref{eq053}, we get \eqref{102}.
\end{proof}

\section{$\sigma$-form of a Painlev\'{e} V Equation}
In this section, by using the Riccati equations and Lemma 3.3, we obtain the second order differential equation satisfied by $\sigma_n(a)$, the logarithmic derivative of Hankel determinant $D_n(a)$ given by \eqref{002}. By introducing the new variable $\tau=2\sqrt{2n}a,\;a\to0^+,\;n\to\infty$, we continue to study the asymptotic behavior of $\sigma_n\left( \tau/(2\sqrt{2n})\right)$ in the variable $\tau$. We show that the limit of $\sigma_n\left( \tau/(2\sqrt{2n})\right)$ in the variable $\tau$ satisfies the $\sigma$-form of a Painlev\'{e} V equation in the variable $\tau$.

Define
\begin{align*}
	\sigma_n(a):=a\dfrac{d}{da}\ln{D_n(a)}.
\end{align*}
From \eqref{Dn} and \eqref{312}, we know that
\begin{align*}
	\sigma_n(a)=a\dfrac{d}{da}\left( \ln{\prod\limits_{j=0}^{n-1}h_j(a)}\right) =-a\sum\limits_{j=0}^{n-1}R_{j}(a).
\end{align*}
Substituting \eqref{310} and \eqref{314} into the above equation to eliminate $a\sum\limits_{j=0}^{n-1}R_{j}(a)$ and $\sum\limits_{j=0}^{n-1}r_{j}(a)$, we get the relationship between $\sigma_n(a)$ and $p(n,a)$ as follows
\begin{align}
	\sigma_n=4p(n,a)+n(n-1)-r_n. \label{eq058}
\end{align}
Plugging (\ref{308}) into (\ref{eq058}) gives 
\begin{align}
	\sigma_n-2a^2r_n-r^2_n=-aR_n(n+r_n)-\dfrac{2ar^2_n}{R_n}.  \label{eq060}
\end{align}
Taking the derivative of (\ref{eq058}) with respect to $a$, and using \eqref{p'} and \eqref{eqrn} to remove $p'(n,a)$ and $r'_n$, we deduce   
\begin{align}
	a\sigma'_n-4a^2r_n=-aR_n(n+r_n)-\dfrac{2ar^2_n}{R_n}. \label{eq061}
\end{align}
Combining (\ref{eq060}) with (\ref{eq061}) leads us to  
\begin{align}
	r^2_n=2a^2r_n+\sigma_n-a\sigma'_n. \label{eq062}
\end{align}
Multiplying \eqref{eqrn} by $a$, we find
\begin{align}
	ar'_n=-aR_n(n+r_n)+\dfrac{2ar^2_n}{R_n}. \label{eq063}
\end{align}
The subtraction and addition of \eqref{eq060} and \eqref{eq063} give us 
\begin{align*}
	ar'_n+\left( \sigma_n-2a^2r_n-r^2_n\right)&=-2aR_n(n+r_n),\\
	ar'_n-\left( \sigma_n-2a^2r_n-r^2_n\right)&=\dfrac{4ar^2_n}{R_n}. 
\end{align*}
Multiplying the above two equations, we are led to  
\begin{align}
	a^2(r'_n)^{2}-(\sigma_n-2a^2r_n-r^2_n)^2=-8a^2r^2_n(n+r_n). \label{eq066}
\end{align}

To obtain our desired second order differential equation for $\sigma_n(a)$, we need to make use of two equations \eqref{eq062} and \eqref{eq066}. From \eqref{eq062}, we get the expressions of $r^2_n(a)$, $r'_n(a)$ and $(r'_n(a))^2$  in terms of $r_n(a)$, $\sigma_n(a)$, $\sigma'_n(a)$ and $\sigma''_n(a)$. Eliminating them associated with $r_n(a)$ in \eqref{eq066}, after simplification, we achieve our goal.
\begin{thm} $\sigma_n(a)$ satisfies the following second order fourth-degree ordinary differential equation
	\begin{equation}
		\begin{aligned}
			&[16a^6-a^2(\sigma''_n)^2-4(a^4+\sigma_n-a\sigma'_n)\left( 4a^2+8n(\sigma_n-a\sigma'_n)-(\sigma'_n)^2\right) ]^2\\
			=&16[4(a^4+\sigma_n-a\sigma'_n)(a\sigma'_n-2\sigma_n)^2+4a^2\sigma''_n(a\sigma'_n-\sigma_n)-a^4(\sigma''_n)^2]\\
			&\cdot[4(2a^2n+\sigma_n)(a^4+\sigma_n-a\sigma'_n)+4a^4-a^2\sigma''_n].\label{sigman}
		\end{aligned}
	\end{equation}
\end{thm}
\begin{proof}
	Plugging (\ref{eq062}) into (\ref{eq066}) to eliminate $r^2_n$, we find 
	\begin{align}
		 (r'_n)^{2}-(\sigma'_n-4ar_n)^2=-8\left[r_n\left( 4a^4+2a^2n+\sigma_n-a\sigma'_n\right) +(2a^2+n)(\sigma_n-a\sigma'_n)\right]. \label{eq067}
	\end{align}
	Differentiating both sides of \eqref{eq062} with respect to $a$ gives us
	\begin{align*}
		r'_n=\dfrac{a(4r_n-\sigma''_n)}{2(r_n-a^2)}.
	\end{align*}
	Squaring both sides of the above equation, and substituting (\ref{eq062}) into the resulting equation to get rid of $r^{2}_n$, we obtain
	\begin{align}
		(r'_n)^2=\dfrac{a^2(4r_n-\sigma''_n)^2}{4(a^4+\sigma_n-a\sigma'_n)}. \label{eq069}
	\end{align}
	Inserting (\ref{eq069}) into (\ref{eq067}) to eliminate $(r'_n)^2$, after simplification, we find
	\begin{align}
		32(a^4+\sigma_n-a\sigma'_n)&(\sigma_n+4a^4+2a^2n)-8a^2\sigma''_n\nonumber\\
		&=\dfrac{1}{r_n}F_n(a)+16a^2r_n[4(a^4+\sigma_n-a\sigma'_n)-1], \label{eq071}
	\end{align}
	where
	$F_n(a)=(a^4+\sigma_n-a\sigma'_n)[4(\sigma'_n)^2-32(2a^2+n)(\sigma_n-a\sigma'_n)]-a^2(\sigma''_n)^2$.
	
	Dividing both sides of (\ref{eq062}) by $r_n$, we have 
	\begin{align*}
		r_n&=2a^2+\dfrac{1}{r_n}(\sigma_n-a\sigma'_n),
	\end{align*}
	which is equivalent to
	\begin{align*}
		\dfrac{1}{r_n}&=\dfrac{r_n-2a^2}{\sigma_n-a\sigma'_n}.
	\end{align*}
	Substituting the above two equations into (\ref{eq071}) to remove $r_n$ and $1/r_n$ respectively, we get the following two equations
	\begin{align*}
		32(a^4+\sigma_n-a\sigma'_n)&(\sigma_n+4a^4+2a^2n)-8a^2\sigma''_n-32a^4[4(a^4+\sigma_n-a\sigma'_n)-1]\nonumber\\
		&=\dfrac{1}{r_n}[F_n(a)+16a^2(\sigma_n-a\sigma'_n)(4(a^4+\sigma_n-a\sigma'_n)-1)],\\
		8(a^4+\sigma_n-a\sigma'_n)&(a\sigma'_n-2\sigma_n)^2+8a^2\sigma''_n(a\sigma'_n-\sigma_n)-2a^4(\sigma''_n)^2\nonumber\\
		&=r_n[F_n(a)+16a^2(\sigma_n-a\sigma'_n)(4(a^4+\sigma_n-a\sigma'_n)-1)].
	\end{align*}
	Multiplying both sides of the above two equations, we arrive at (\ref{sigman}). 
\end{proof}

To continue, we assume $n\to\infty$ and $a\to0^+$ such that 
\begin{align}
	\tau:=2\sqrt{2n}a\label{eq079}
\end{align}
is fixed. For the explanation of this double scaling, see \cite{29, 20}. Define
\begin{align}
	\sigma(\tau)&:=\lim_{n \to \infty} \sigma_n\left( \dfrac{\tau}{2\sqrt{2n}}\right).\label{eq080}
\end{align}

Solving $r_n$ from \eqref{eq062}, we find
\begin{align*}
	r_n\left(a\right)=a^2\pm\sqrt{a^4+\sigma_n\left(a\right)-a\sigma'_n\left(a\right)}. 
\end{align*}
Substituting (\ref{eq079}) and (\ref{eq080}) into the above equation, we have 
\begin{align*}
	r_n\left( \dfrac{\tau}{2\sqrt{2n}}\right)=\dfrac{\tau^2}{8n}\pm\sqrt{\dfrac{\tau^4}{64n^2}+\sigma_n\left( \dfrac{\tau}{2\sqrt{2n}}\right)-\tau\dfrac{d}{d\tau}\sigma_n\left( \dfrac{\tau}{2\sqrt{2n}}\right)}. 
\end{align*}
We find that the limit of $r_n\left( \tau/(2\sqrt{2n})\right)$ is 
\begin{align*}
	\lim_{n \to \infty} r_n\left( \dfrac{\tau}{2\sqrt{2n}}\right)=\pm\sqrt{\sigma(\tau)-\tau\sigma'(\tau)}.
\end{align*}
Hence, we conclude that $r_n\left( \tau/(2\sqrt{2n})\right) \sim O(1)$ as $n\to\infty$ and we thus define
\begin{align}
	r(\tau):=\lim_{n \to \infty} r_n\left( \dfrac{\tau}{2\sqrt{2n}}\right).\label{eq0001}
\end{align}

Similarly, using \eqref{320} and with the aid of \eqref{eq079} and \eqref{eq0001}, we know that 
\begin{align*}
	R_n\left( \dfrac{\tau}{2\sqrt{2n}}\right)=&
	\pm\dfrac{\sqrt{2n\left\lbrace \left[\dfrac{d}{d\tau}r_n\left( \dfrac{\tau}{2\sqrt{2n}}\right) \right]^2+r^2_n\left( \dfrac{\tau}{2\sqrt{2n}}\right) \right\rbrace +2r^3_n\left( \dfrac{\tau}{2\sqrt{2n}}\right) }}{n+r_n\left( \dfrac{\tau}{2\sqrt{2n}}\right)}\\
	&-\dfrac{\sqrt{2n}\cdot\dfrac{d}{d\tau}r_n\left( \dfrac{\tau}{2\sqrt{2n}}\right)}{ n+r_n\left( \dfrac{\tau}{2\sqrt{2n}}\right)},
\end{align*}
which indicates that $R_n\left( \tau/(2\sqrt{2n})\right)\sim O(n^{-\frac{1}{2}})$ as $n\to\infty$. Thus, we define 
\begin{align}
	R(\tau):=\lim_{n \to \infty} \sqrt{n}R_n\left( \dfrac{\tau}{2\sqrt{2n}}\right).\label{eq0002}
\end{align}

Using the definitions \eqref{eq080}-\eqref{eq0002}, we obtain the second order differential equations satisfied by $R(\tau)$, $r(\tau)$ and $\sigma(\tau)$ from \eqref{eq048}, \eqref{eq053} and \eqref{sigman}.
\begin{thm}
	$r(\tau)$ and $R(\tau)$ defined by \eqref{eq0001}-\eqref{eq0002} satisfy the following two second order differential equations 
	\begin{align}
		R''=\left(\dfrac{1}{2R}+\dfrac{\sqrt{2}}{2(\sqrt{2}R-\tau)}\right)(R')^2-&\left(\dfrac{1}{\tau}+\dfrac{1}{\sqrt{2}R-\tau}\right)R'+\dfrac{\sqrt{2}}{2\tau}R^2-\dfrac{R}{2},\label{eq0004}\\
		4r^2((r')^2+r^2)-&\tau^2(r''+r)^2=0.\label{eq0003}
	\end{align}
\end{thm}
\begin{proof}
	Replacing $a$ by $\tau/(2\sqrt{2n})$ in \eqref{eq048}, and dividing its both sides by $8\sqrt{n}$, we obtain
	\begin{align*}
		\dfrac{d^2}{d\tau^2}\sqrt{n}R_n\left(\dfrac{\tau}{2\sqrt{2n}}\right)
		=\dfrac{\tau-2\sqrt{2n}R_n\left( \dfrac{\tau}{2\sqrt{2n}}\right)}{2\sqrt{n}R_n\left( \dfrac{\tau}{2\sqrt{2n}}\right)\left[ \tau-\sqrt{2n}R_n\left( \dfrac{\tau}{2\sqrt{2n}}\right)\right]}\left[ \dfrac{d}{d\tau}\sqrt{n}R_n\left( \dfrac{\tau}{2\sqrt{2n}}\right)\right]^2
	\end{align*}
	\begin{align*}
		&-\dfrac{\sqrt{2n}R_n\left( \dfrac{\tau}{2\sqrt{2n}}\right)}{\tau\left[ \tau-\sqrt{2n}R_n\left( \dfrac{\tau}{2\sqrt{2n}}\right)\right]}\dfrac{d}{d\tau}\sqrt{n}R_n\left( \dfrac{\tau}{2\sqrt{2n}}\right)\\
		&+\dfrac{\sqrt{n}}{4\tau}R_n\left( \dfrac{\tau}{2\sqrt{2n}}\right)\left[ \tau-\sqrt{2n
		}R_n\left( \dfrac{\tau}{2\sqrt{2n}}\right)\right]\left[ -2-\dfrac{1}{n}+\dfrac{\tau^2}{8n^2}-\dfrac{\tau \sqrt{n}R_n\left( \dfrac{\tau}{2\sqrt{2n}}\right)}{2\sqrt{2}n^2}\right].
	\end{align*}
	As $n \to \infty$, in view of \eqref{eq0002}, we come to \eqref{eq0004}.
	
	Similarly, replacing $a$ by $\tau/(2\sqrt{2n})$ in (\ref{eq053}), we have  
	\begin{align*}
		16\left[\dfrac{\tau^2}{8n}+r_n\left( \dfrac{\tau}{2\sqrt{2n}}\right) \right]^2&\left\lbrace  n\left[ \left( \dfrac{d}{d\tau}r_n\left( \dfrac{\tau}{2\sqrt{2n}}\right)\right) ^2+r_n^2\left( \dfrac{\tau}{2\sqrt{2n}}\right)\right] +r_n^3\left( \dfrac{\tau}{2\sqrt{2n}}\right) \right\rbrace  \\
		&-\dfrac{\tau^2}{n}\left\lbrace  2n\left[  \dfrac{d^2}{d\tau^2}r_n\left( \dfrac{\tau}{2\sqrt{2n}}\right)+r_n\left( \dfrac{\tau}{2\sqrt{2n}}\right)\right] +3r^2_n\left( \dfrac{\tau}{2\sqrt{2n}}\right) \right\rbrace ^2=0.
	\end{align*}
	Dividing both sides of the above equation by $n$, we find
	\begin{align*}
		16\left[\dfrac{\tau^2}{8n}+r_n\left( \dfrac{\tau}{2\sqrt{2n}}\right) \right]^2&\left\lbrace \left( \dfrac{d}{d\tau}r_n\left( \dfrac{\tau}{2\sqrt{2n}}\right)\right) ^2+r_n^2\left( \dfrac{\tau}{2\sqrt{2n}}\right)+\dfrac{1}{n}r_n^3\left( \dfrac{\tau}{2\sqrt{2n}}\right) \right\rbrace  \\
		&-\tau^2\left\lbrace 2\left[  \dfrac{d^2}{d\tau^2}r_n\left( \dfrac{\tau}{2\sqrt{2n}}\right)+r_n\left( \dfrac{\tau}{2\sqrt{2n}}\right)\right] +\dfrac{3}{n}r^2_n\left( \dfrac{\tau}{2\sqrt{2n}}\right) \right\rbrace ^2=0.
	\end{align*}
	As $n\to \infty$, in view of \eqref{eq0001}, we are led to (\ref{eq0003}).
\end{proof}

Finally, substituting $a=\tau/(2\sqrt{2n})$ into \eqref{sigman}, and sending $n$ to $\infty$, we obtain the following second order differential equation for $\sigma(\tau)$.
\begin{thm}
	$\sigma(\tau)$ satisfies the following second order differential equation
	\begin{align}
		(\tau\sigma'')^2=-4(\sigma-\tau\sigma'-(\sigma')^2)(\sigma-\tau\sigma'),\label{eq081}
	\end{align}
	which is identified to be the Jimbo-Miwa-Okamoto $\sigma$-form of a Painlev\'{e} V equation \cite{17, 18} with $v_0=v_1=v_2=v_3=0$.
\end{thm}
\begin{remark}
	 Recall our weight function $w(x;a)=\e^{-x^2}\left( A+B\theta(x^2-a^2)\right) ,\;A\ge0 ,\; A+B\ge0$. When $A=0$ and $B=1$, the associated Hankel determinant $D_n(a)$ reads
	 \begin{align*}
	 	D_n(a)=\mathrm{det}\left(\int_{R\setminus [-a,a]}x^{i+j}\e^{-x^2}\theta(x^2-a^2)
	 	dx\right)_{i,j=0}^{n-1},
	 \end{align*}
	 which was studied in \cite{20}. Our equations satisfied by the logarithmic derivative of the above Hankel determinant $D_n(a)$ for finite $n$ and large $n$, i.e. \eqref{sigman} and \eqref{eq081}, agree with (2.3) and (2.4) of \cite{20}.
\end{remark}

\section*{Acknowledgements}
This work was supported by National Natural Science Foundation of China under grant	numbers 12101343, by Shandong Provincial Natural Science Foundation
with project number ZR2021QA061, and by Scientific Research Fund for Talented Scholars of Qilu University of Technology (Shandong Academy of Sciences) under grant number 2023RCKY245.

\end{document}